\documentclass[aps,pra,onecolumn,nofootinbib,11pt,superscriptaddress]{revtex4-2}

\usepackage[utf8]{inputenc}
\usepackage[margin=1in]{geometry} 
\usepackage{amsmath, amssymb, amsfonts}
\usepackage{graphicx}
\usepackage{physics}
\usepackage[ruled,vlined]{algorithm2e}
\usepackage{hyperref} 
\hypersetup{colorlinks=true, linkcolor=blue, citecolor=blue, urlcolor=blue}

\begin{document}

\title{Homodyne Detection of Temporally Resolved Quantum States}

\author{Owen Sandner}
\author{Brendan Mackey}
\author{Yuyang Liu}
\affiliation{Department of Physics and Astronomy, University of Victoria, Victoria, BC, Canada}

\author{Connor Kupchak}
\affiliation{Department of Electronics, Carleton University, Ottawa, ON, Canada}

\author{Andrew MacRae}
\email{macrae@uvic.ca}
\affiliation{Department of Physics and Astronomy, University of Victoria, Victoria, BC, Canada}

\begin{abstract}
We present an analysis of the time domain measurement of temporally resolvable quantum states using balanced homodyne detection. Our approach outlines a formalism of detecting quantum states in arbitrary temporal modes via projection of the temporal mode onto a natural detector basis. We then present an algorithm for simulating the resultant photocurrent of continuous homodyne detection in the presence of a temporally resolved mode, and use this algorithm to explore the effects of realistic measurement errors on marginal reconstruction and quantum state tomography. A complete implementation of the method is provided through open source code on a GitHub repository.
\end{abstract}
\maketitle

\section{Introduction}
Balanced Homodyne Detection (BHD) has become a ubiquitous tool in quantum optics, quantum communication, and optical quantum computing. By combining a quantum optical state with a strong local oscillator on a symmetric (50/50) beamsplitter, and subtracting the two resultant photocurrents, BHD coherently amplifies fluctuations of the quantum optical state while canceling any technical noise in the system. Since its inception~\cite{yuen1983noise,walker1984simultaneous}, BHD has been crucial in measurement of quadrature squeezing~\cite{slusher1985observation}, quantum state tomography~\cite{lvovsky2009continuous}, and more recently, continuous variable-based logical qubits~\cite{weedbrook2012gaussian}.

As a phase sensitive amplifier, the homodyne detector selects a particular spatio-temporal mode of the quantum optical field corresponding to that of the local oscillator~\cite{leonhardt1997measuring}. For quantum states generated by nonlinear crystals with phasematching bandwidth much wider than the electronics of the detection system, the temporal mode of the local oscillator is shaped to optimize the overlap with the quantum state either with \textit{a priori} knowledge~\cite{ourjoumtsev2006generating} or adaptively~\cite{polycarpou2012adaptive} and the lower bandwidth detector integrates each pulse into a measurement in this mode~\cite{kumar2012versatile}. 

For narrowband states with spectral width less than the detector bandwidth, the state can be temporally resolved and in the limit of large detector bandwidth, continuously measure the quantum state. In this case, a CW local oscillator may be used and the instantaneous photocurrent can be integrated against a mode weighting function to reproduce the state~\cite{macrae2012tomography}. For reconstructing the quantum state from arbitrary processes, it is therefore essential to have knowledge of the spatio-temporal mode in which the quantum state exists.

The relationship between modes and quantum states has been explored since the birth of quantum optics, with the seminal work of Glauber and colleagues~\cite{titulaer1966density}. Recent reviews for temporal modes~\cite{raymer2020temporal} and spatio-temporal modes~\cite{fabre2020modes} have laid out this relationship explicitly. BHD has been used directly to measure the complex spatial mode of a quantum state by observing the autocorrelation of the photocurrent~\cite{qin2015complete} and more general methods have been developed more recently~\cite{morin2020accurate, takase2019complete, huo2020direct}. In contrast, this work treats the time dependent photocurrent as a basis vector in a modal Hilbert space and treats mode mismatch as a geometric projection onto a larger dimensional space rather than as a generalized loss parameter. This is of particular interest to recently developed quantum information protocols which use homodyne detection for continuous variable quantum computing via Gottesman-Kitaev-Preskill (GKP) states. 

Our paper is outlined as follows: In Section~\ref{sec:ctsdet}, we outline the general theory of temporal modes in quantum optics and give a brief review of homodyne theory. We define two pictures of time resolved homodyne detection corresponding to the time-bin, or ``detector basis'' and the ``principal basis'' define through the mode of the quantum state. Next, in Section~\ref{sec:algo}, we outline an algorithm to simulate realistic homodyne photocurrent in the presence of a time resolved quantum state, and show how mode overlap affects the reconstructed mode. Following this, in Section~\ref{sec:apps} we demonstrate the utility of this algorithm for quantifying errors in the reconstruction of quantum states in the presence of various experimental imperfections. Finally, Section~\ref{sec:conc} summarizes this work and discusses extensions and further applications of this work.

\section{Continuous Detection of Temporal Modes}
\label{sec:ctsdet}

\subsection{Temporal Modes in Quantum Optics}
In classical electromagnetism, modes are defined by solutions to the Maxwell equations satisfying the boundary conditions imposed by a physical system. Each individual solution is a mode and the complete set of all orthonormal solutions forms a modal basis. In free space, the most natural basis is that of monochromatic plane waves, given as:

\begin{equation}
    \label{eq:def_planemode}
    \mathbf{u}_{\mathbf{k},\hat{\epsilon}}(\mathbf{r},t) = \hat{\epsilon} e^{i(\textbf{k}\cdot\textbf{r}-\omega t)}
\end{equation}

\noindent for a polarization vector $\hat{\epsilon}$ and a wave-vector $\mathbf{k} = \frac{\omega}{c}\hat{k}$. It should be noted that plane waves are non-physical as they are non-normalizable. However, they are a convenient basis for physical modes akin to Fourier analysis of other physical systems. Plane waves form the basis for formal quantization, in which the electric field operator is described in terms of bosonic operators $\hat{a}_{\mathbf{k},\hat{\epsilon}}$ corresponding to the annihilation of a photon in the plane wave $(\mathbf{k},\hat{\epsilon})$\cite{loudon2000quantum}.

In quantum optics, the quantum state of light can be described in terms of Fock states $\ket{n}$ representing $n$ photon excitations of the field. For a pure state, we can write this generally as:

\begin{equation}
    \label{eq:def_single_mode_q_state}
    \ket{\psi} =\sum_{n=0}^\infty c_n\ket{n}
\end{equation}

\noindent whereas a mixed state can be represented as a density matrix $\hat{\rho} = \sum_k\Pr_k\ket{\psi_k}\bra{\psi_k}$.

In writing equation~\ref{eq:def_single_mode_q_state}, it is implied that the state exists in the wavefunction in a particular mode. Specifically, \textit{modes act as containers into which quantum states may be placed}~\cite{raymer2020temporal}. Additionally, owing to the relation between optics (modes) and quantum mechanics (states), quantum optics necessitates a relation between the mode and the quantum state which occupies it. 

Note that while equation~\ref{eq:def_single_mode_q_state} specifies the quantum state of a single mode, it is not necessarily a complete description of the quantum state. More generally, for a mode set $\left\{ f_m\right\}$, we can write the full state as a tensor product of terms defined in equation~\ref{eq:def_single_mode_q_state}:

\begin{equation}
    \label{eq:def_multi_mode_q_state}
    \ket\psi = \ket{\psi_1}\ket{\psi_2}\cdots\ket{\psi_N}
\end{equation}

\noindent where $\ket{\psi_m}$ refers to the state in mode $f_m$. States for which there exists a mode basis for which all states but one are in the vacuum state, $\ket\psi_{sm} = \ket\psi\ket0\ket0\cdots\ket0$, are said to be \textit{intrinsically single-mode}. States for which there exists no such basis are \textit{intrinsically multi-mode}.

A common scenario in optics is where a narrowband field propagates in a beam-like geometry in a slowly varying amplitude envelope. In this setting, it is convenient to separate the spatial mode into a diffractive \textit{transverse mode} and a longitudinal \textit{temporal mode}\cite{raymer2020temporal}. In this work we focus solely on temporal modes, thus we consider a fixed polarization and take the propagation direction to be $\hat{z}$. This reduces the temporal mode to a scalar field labeled by its wave-number $k = \omega/c$ or equivalently, by its frequency. We then define $\hat{a}(\omega)$ as the annihilation operator for a given plane wave with frequency $\omega$ such that

\begin{equation}
    \label{eq:monochromatic_commutation}
    \left[\hat{a}(\omega),\hat{a}^\dagger(\omega^\prime)\right] = \delta(\omega-\omega^\prime)
\end{equation}

\noindent and we write the field in terms of its positive frequency component as ~\cite{glauber1963coherent}:

\begin{equation}
    \label{eq:plane_wave_continuum}\hat{E}^{(+)}(z,t) = i\int_0^\infty d\omega\sqrt{\frac{\hbar\omega}{2\varepsilon_0}}\hat{a}(\omega)e^{i\frac{\omega}{c}(z-ct)}.
\end{equation}

Equation~\ref{eq:plane_wave_continuum} represents the field in terms of the continuous basis of plane wave modes. However, we are not restricted to plane wave modes. Limiting our analysis to a finite temporal extent, we can discretize the description. Starting with some discrete orthonormal mode basis $\left\{f_m(t)\right\}$ with Fourier transforms $\left\{\tilde{f}_m(\omega)\right\}$, we can consider a weighted distribution of these annihilation operators for a particular mode, thus defining the annihilation operator for the discrete mode  $f_m(t)$:

\begin{equation}
\label{eq:single_phot_in_mode}
    \hat{b}_m\equiv\int\tilde{f}_m(\omega)\hat{a}(\omega)d\omega.
\end{equation}

\noindent In this way we have a discrete set of quantum modes, expressed in terms of the continuous plane wave operators $\hat{a}(\omega)$. Using this we can write equation~\ref{eq:plane_wave_continuum} in terms of discrete components as:

\begin{equation}
        \label{eq:discrete_field}
    \hat{E}^{(+)}(z,t) = \sum_kf_k(z,t)\hat{b}_k.
\end{equation}

It is important to note that these modes obey the discrete equivalent of the commutation relation in equation~\ref{eq:monochromatic_commutation}:

\begin{equation}
    \label{eq:discrete_commutation}
    \left[\hat{b}_j,\hat{b}^\dagger_k\right]= \delta_{jk}.
\end{equation}

\noindent We can then apply the pre-existing formalism for single mode quantum optics to each of these mode elements. Arbitrary pure states can be written in terms of the series expansion:

$$
\ket\psi_k = \sum_{n=0}^\infty c_n\ket{n}_k = \sum_{n=0}^\infty\frac{c_n}{\sqrt{n!}}\left(\hat{b}_k^\dagger\right)^n\ket{\mathrm{vac}}.
$$

\noindent For example, we can write a squeezed state in mode $f_k$:

\begin{subequations}

\begin{equation}
    \ket{\xi}_k = \exp\left[\xi(\hat{b}_k^\dagger)^2 -\xi^*\hat{b}_k^2 \right]\ket{\mathrm{vac}}
\end{equation}

\noindent for squeezing parameter $\xi$, or a coherent state as

\begin{equation}
    \ket{\alpha}_k = \hat{D}(\alpha)\ket{\mathrm{vac}} = \exp\left[\alpha\hat{b}^\dagger_k-\alpha^*\hat{b}_k \right]\ket{\mathrm{vac}}
\end{equation}
\end{subequations}

Fabre and Treps (2020) introduced the concept ``modal space'' which is a Hilbert space of possible modes \cite{fabre2020modes}. In this formalism we can consider a column vector of bosonic operators representing all possible modes:

\begin{equation}
    \vec{\hat{b}}^\dagger\leftrightarrow\left\{\hat{b}_k^\dagger\right\}    
\end{equation}

\noindent The interplay between modes and states is epitomized by the relation between single photons in two separate modes $f$ and $g$ from equation~\ref{eq:single_phot_in_mode}~\cite{fabre2020modes}:
\begin{equation}
    \label{eq:modestaterelation}
    \langle1_f\vert 1_g\rangle = \mathbf{f}^{*}\cdot\mathbf{g}
\end{equation}

\noindent We can then perform a change of basis of our quantum state by a unitary operation:

\begin{equation}
    \vec{\hat{c}}^\dagger = \hat{U}\vec{\hat{b}}^\dagger
\end{equation}

\noindent As long as the transformation $\hat{U}$ is unitary, all of the standard bosonic relations such as the number $\hat{n}$, position $\hat{X}$, and momentum $\hat{P}$ operators hold:

\begin{subequations}
\label{eq:qo_in_new_basis}
    \begin{align}
        \hat{n}_c &= \hat{c}^\dagger\hat{c} \\
        \hat{X}_c &= \frac{\hat{c}+\hat{c}^\dagger}{\sqrt{2}} \\
        \hat{P}_c &= \frac{\hat{c}-\hat{c}^\dagger}{\sqrt{2}i} \\
    \end{align}
\end{subequations}

We are thus free to choose any basis for quantum optics, though some bases may be more natural than others. A common choice in free space is the \textit{time-bin} basis, which is especially relevant for continuous photodetection. A mode in this basis is defined by discretizing time into rectangular bins of width $\Delta t$:

\begin{equation}
    f_k(t) = 
\begin{cases}
    1,& \text{if } t\in [t_k,t_k+\Delta t) \\
    0,              & \text{otherwise}
\end{cases}
\end{equation}

In such a basis, many modes may be occupied with vacuum and non-vacuum states and thus the state may appear to be multi-mode. However, often one can find a \textit{principal basis} for which there are a minimal number of modes occupied by the quantum state. For temporal modes, this may be completely determined by measuring the auto-correlation matrix of the homodyne photocurrent $\left\langle I(t_i)I(t_j)\right\rangle_{ij}$ across all time bins for several local oscillator frequencies~\cite{qin2015complete} (see section~\ref{sec:bhd-theo} below.) States whose autocorrelation matrices have a single non-zero eigenvalue are intrinsically single-mode with the principal mode defined by the corresponding eigenvector.

Although the choice of time-bin widths, and thus a particular detector mode basis is somewhat arbitrary, the effect of discretization of the temporal mode is negligible whenever the variation of the temporal mode is much slower than both the time response of the detector and the time bin spacing. Equivalently, we require the quantum state to be narrow-band with respect to both the electronics and the sampling rate of the detection system.

\subsection{Balanced Homodyne Detection}
\label{sec:bhd-theo}
In balanced homodyne detection~\cite{yuen1983noise}, a quantum optical state $\hat\rho$ in a particular spatial-temporal mode $f_\rho(z,t)$ is overlapped with strong ``local oscillator'' field in mode $f_{LO}(z,t)$, usually assumed to be a coherent state $\ket{\alpha}$ with mean photon number $\overline{n}\gg1$, on a symmetric beam-splitter. The two outputs are sent to matched photodetectors and the photocurrents from each are subtracted to form the difference current $\hat{i}_{HD}(t) = \hat{i}_a(t)-\hat{i}_b(t)$. If the quantum and local oscillator modes are matched, $f_\rho = f_{LO} \equiv f$, the difference in these photocurrents can be shown to be directly proportional to a measurement of the generalized quadrature of the quantum state~\cite{kumar2012versatile}:

\begin{equation}
    \label{eq:bhd_photocurrent}
    \hat{i}_{HD}(t)\propto\hat{q}_{f}^\theta(t)
\end{equation}
where, $\theta$ is the relative phase between the quantum state with respect to the local oscillator and the generalized quadrature for mode $f$ with bosonic operators $\hat{b}$ is defined as:

\begin{equation}
    \label{eq:gen_quad}
    \hat{q}_{f}^\theta(t) = \frac{\hat{b}_f(t)e^{i\theta} + \hat{b}_f(t)^\dagger e^{-i\theta}}{\sqrt{2}} = \hat{X}_f(t)\cos\theta - \hat{P}_f(t)\sin\theta.
\end{equation}
We thus see that by scanning the phase of the local oscillator, we can measure any angle in quasi-phase space $(\hat{X},\hat{P})$. 

In the case that the modes of the quantum state and the local oscillator are not matched, the balanced homodyne detector performs a projective measurement of the quantum state into the mode of the local oscillator. This mode-mismatch is equivalent to a mixture with vacuum and can be treated as an effective loss at a beam-splitter with transmittivity $\eta$~\cite{kumar2012versatile}.

If the quantum state is generated at a fixed time with a well defined spatiotemporal mode, such as ultra-fast pulsed single photon states in parametric down-conversion~\cite{lvovsky2001quantum}, the local oscillator can be adjusted adaptively to maximize the overlap with the mode of the quantum state~\cite{polycarpou2012adaptive}. If continuous detection is employed, such as in heralded single photon generation from a continuously driven system~\cite{macrae2012tomography}, the quadrature must be inferred from the photocurrent of the continuous-wave local oscillator. The fact that the local oscillator is continuous and the quantum state is a temporally varying mode can be seen as mode-mismatch: the local oscillator is naturally measuring in the time-bin basis which we shall call the  ``detector basis'' whereas the quantum state occupies a single ``principal mode''. Continuous homodyne detection thus projects the principal mode into the detector basis.

To describe this formally, consider the analysis from the previous section. Assuming for now the measurement of a pure state $\ket{\psi}$ in mode $g_1(z,t)$, we can define an orthonormal basis $\{g_k(z,t)\}$ via the Gram Schmidt procedure such that all other modes $g_{k>1}$ contain the vacuum state $\ket{0}_k$:

\begin{subequations}
\begin{equation}
    \label{eq:init_state}
    \ket\psi_g^{\mathrm{tot}} = \ket\psi_1\ket0_2\ket0_3\ket0_4\cdots
\end{equation}
Changing to a new basis set $\{h_k(z,t)\}$ via a unitary operation will subsequently express each state as a linear combination $h_k = \sum_j\gamma_{jk}f_j$ and so each mode now contains a new state $\ket\phi$ which is a combination of vacuum and $\ket\psi$:

\begin{equation}
    \label{eq:init_state_newbasis}
    \ket\psi_h^{\mathrm{tot}} = \ket\phi_1\ket\phi_2\ket\phi_3\ket\phi_4\cdots
\end{equation}
\end{subequations}

\noindent Measuring $h_k$ in the new basis will thus correspond to a partial measurement of the state with a mixture of vacuum. 

\begin{figure}
    \centering
    \includegraphics[width=0.7\linewidth]{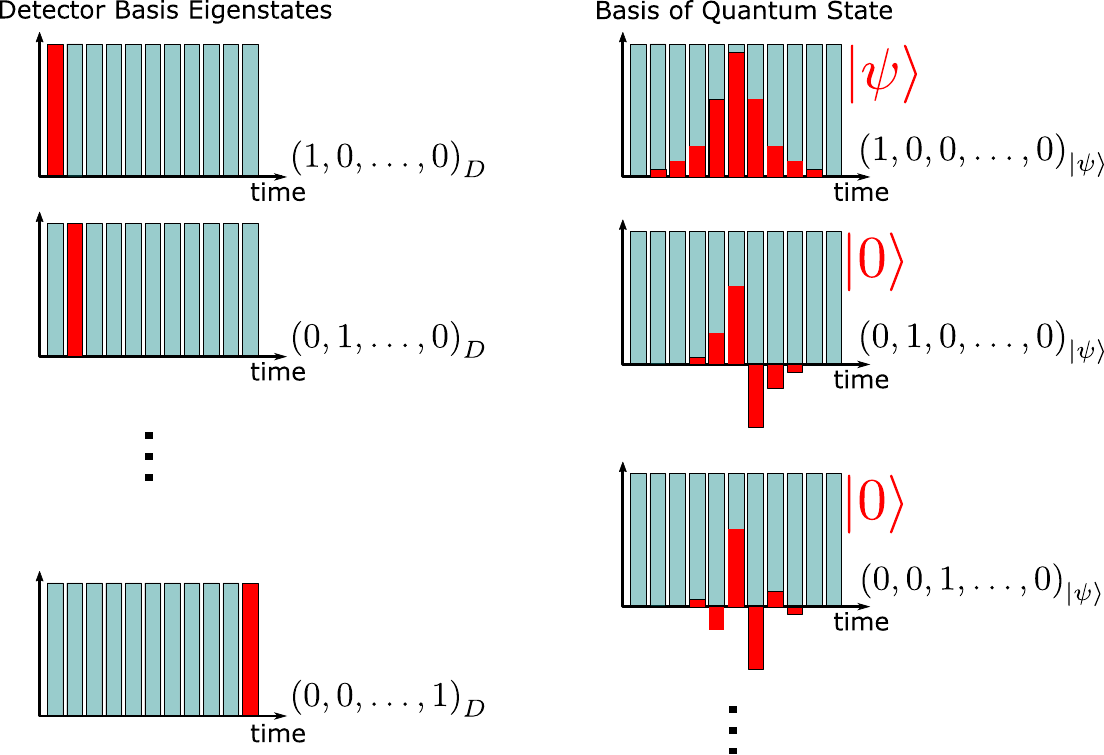}
    \caption{A quantum state exists in a particular mode which is an element of a basis $\{f\}_k$ (right). The detector naturally measures in time-bin mode (left). In the detector basis, the state is spread over multiple time bins, as are the additional vacuum modes.}
    \label{fig:mode_cartoon}
\end{figure}

In light of this, consider continuous homodyne detection of a quantum state $\hat{\rho}$ with a continuous local oscillator $\ket\alpha_{LO}$. The detector photocurrent is naturally measuring in the time bin basis, with each sample representing a time bin dictated by the local oscillator and the detector response. BHD projects the state into this basis so that each time bin represents a mixture of the quantum state and the vacuum. We can also represent the detector photocurrent in the basis of the quantum state. In this basis both the quantum state \textbf{and} the vacuum states are mixtures of time bin modes. The total photocurrent is the sum of all of these modes. This is depicted in Figure~\ref{fig:mode_cartoon}. The modes $\{f_k(z,t)\}$ projected in the detector basis $\{h_k(z,t)\}$ exhibit orthonormality, we can completely recover the state in a given mode by taking the inner product with the continuous homodyne current such that up to a scaling constant:

\begin{equation}
    \label{eq:overlap_integral}    
    q_{\psi,k} = \langle f(t)\vert i(t)\rangle = \sum_jf_{j,k}g_j \approx\int f_k(t)i(t)dt.
\end{equation}

\section{Algorithm for Simulating Continuous Homodyne Measurements}
\label{sec:algo}

As described in \ref{sec:bhd-theo}, a balanced homodyne detector performs a quadrature measurement of the quantum state, with phase $\theta$ defined by the local oscillator. This measurement produces a value sampled from the corresponding marginal distribution $\Pr(q_\theta)$. A quantum state $\hat{\rho}$, along with its marginal distributions $\Pr(q_\theta)$, is completely specified by its Wigner function~\cite{leonhardt1997measuring}:

\begin{equation}
    \label{eq:def_wigfunc}
W(x,p) = \frac{1}{2\pi}\int_{-\infty}^\infty e^{ipq}\left\langle x-\frac{q}{2}\vert\hat{\rho}\vert x + \frac{q}{2}\right\rangle dq.
\end{equation}

\noindent Specifically, for a particular rotation in phase-space $\theta$, the marginal probability distribution $\Pr(q_\theta)$ is realized via integration of the Wigner function along $q_\theta$, accomplished by rotating by $-\theta$ and integrating along $p$:

\begin{equation}
\Pr(q_\theta) = \int_{-\infty}^\infty W(x\cos\theta - p\sin\theta,x\sin\theta +p\cos\theta)dp.
\end{equation}

Efficient methods exist to calculate the Wigner function~\cite{qutip5}, allowing us to determine its marginal and hence its cumulative distribution function (CDF)~\cite{RadonJohann1986}. Such methods allow us to generate a quadrature sample from any quantum state $\hat{\rho}$ using a numerical sampling method. One such method is inverse transform sampling (ITS)~\cite{devroye2006nonuniform}. ITS generates a uniform random sample $x\in[0,1]$ and evaluates the inverse of a given distribution at $x$. Since the inverse of a probability distribution may not be well-known, an approximate method is to use the cumulative distribution $g(x)$ of $Pr(x)$ and interpolate its inverse at $x$. This is also computationally efficient as the CDF is monotonic and thus has a well-defined inverse, which we can find numerically by simply swapping the $x$ and $g(x)$ arrays in an interpolation function.

Given a mode $f_0(t)$ containing our quantum state, we can express $f_0(t)$ in the detector basis and generate from it an orthonormal basis using the Gram-Schmidt procedure or QR factorization \cite{axler2024linear}. In measurement, $f_0$ is weighted by a sample from the quantum state marginal along $\theta$ and the remaining modes are weighted by samples from the background state distribution. Note that often this background is taken to be vacuum which we denote $\ket{\text{vac}}$, but could be any state, such as a thermal distribution. The sum of all weighted modes form a simulation of a single shot of continuous BHD photocurrent. Our algorithm for implementing the above description is outlined below:

\begin{algorithm}[H]
\label{alg:simhomodyne}
\caption{Time-Resolved Homodyne Simulation}
\SetAlgoLined
\KwIn{Quantum state $\hat \rho$, temporal mode $f_0(t)$, number of bins $N$, background state $\ket{\text{vac}}$, gain coefficient $g$}
\KwOut{Photocurrent $i(t_k)$}

Generate orthonormal basis $\{f_0, f_1, \ldots, f_{N-1}\}$ via QR decomposition\;
\For{$j \gets 0$ \KwTo $N-1$}{
    \eIf{$j = 0$}{
        Sample $x_j$ from quadrature marginal of $\hat \rho$\;
    }{
        Sample $x_j$ from background distribution $\ket{\text{vac}}$ \;
    }
}
Compute $i(t_k) \gets g \sum_j x_j f_j(t_k)$\;
\Return{$i(t_k)$}\;
\end{algorithm}

\begin{figure}
    \centering
    \includegraphics[width=0.75\linewidth]{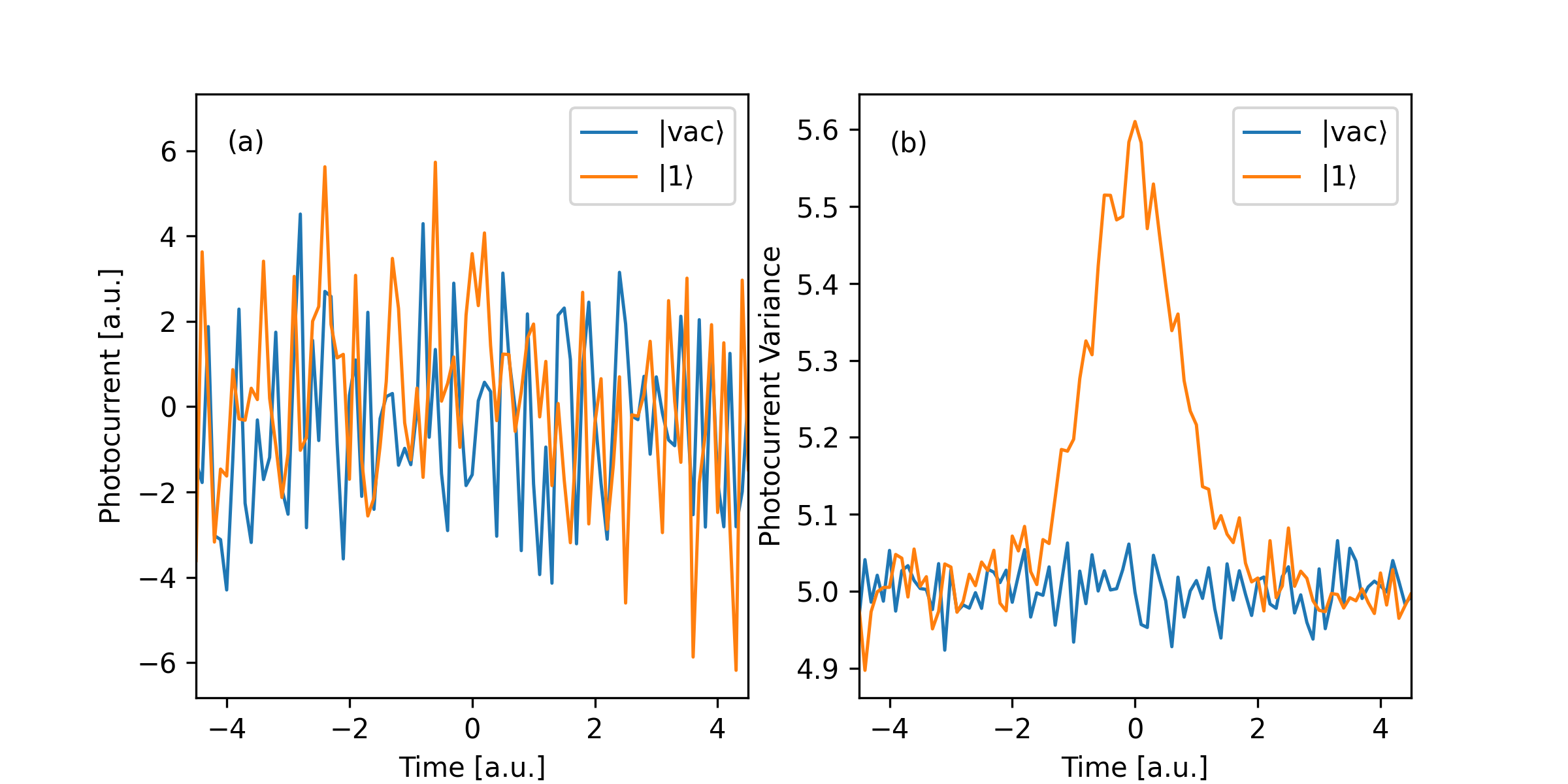}
    \caption{(a) Example photocurrent for a single simulated trace of both vacuum and a Fock state $\ket{1}$. The single photon and vacuum cases are nearly indistinguishable for a single trace as the photon's quadrature measurement occupies a single mode within this trace. This is seen in (b) where the variance for each time bin over $5\times10^4$ traces is shown and the higher variance photon begins to appear.}
    \label{fig:pw_var}
\end{figure}

\begin{figure}
    \centering
    \includegraphics[width=0.75\linewidth]{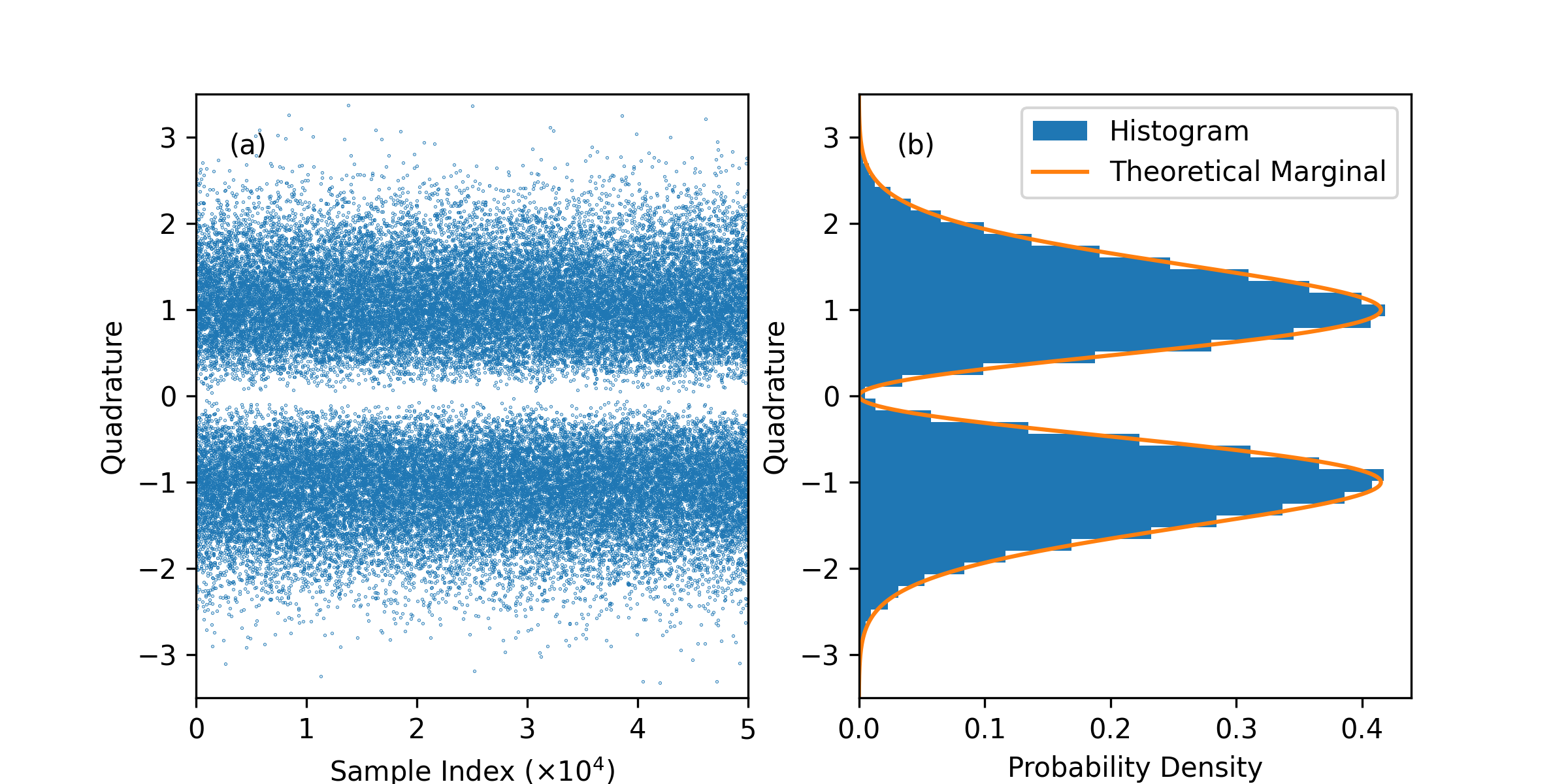}
    \caption{
    (a) $5\times10^4$ integrated homodyne traces, giving the corresponding quadrature sample. (b) Histogram of these quadrature samples, revealing the single-photon Fock state marginal being measured by the detector.}
    \label{fig:sing_phot_marg}
\end{figure}

A sample trace of this process is shown for a single photon Fock state in a Gaussian temporal mode in figure~\ref{fig:pw_var}. The photon's quadrature measurement produces a single number weighting a particular mode of the photocurrent. However, the variance of a single photon's marginal is three times that of the vacuum, therefore evaluation of the pointwise variance over the trace will signal the presence of a single photon. Furthermore, by integrating the photocurrent weighted by state's principal mode, we recover the quadrature sample, and by taking many such measurements, we may recover the marginal distribution as is apparent in figure~\ref{fig:sing_phot_marg}.

\section{Application to Measurement Under Realistic Experimental Conditions}
\label{sec:apps}

To demonstrate the utility of our technique, we quantify the effect of various measurement errors common in BHD. We first introduce a similarity measure known as \textit{Bhattacharyya Coefficient} appropriate for quantifying the similarity of two marginals. We then quantify the impacts of modal mismatch, timing jitter, and phase jitter on BHD measurement.

\subsection{Bhattacharyya Coefficient}
\label{subsec:B}
As BHD samples a marginal distribution rather than directly measuring from the Wigner function of a state, we use the Bhattacharyya coefficient \cite{kang2015bhattacharyya},
\begin{equation}\label{eq:bhat_coeff}
    B({\Pr}_1(t),{\Pr}_2(t)) = \int_{-\infty}^\infty \sqrt{{\Pr}_2(t){\Pr}_2(t)}dt,
\end{equation}
which is a standard measurement of similarity between Probability Density Functions (PDFs). This coefficient satisfies $0\le B\le 1$ for two PDFs ${\Pr}_1(t)$ and ${\Pr}_2(t)$, with $B=1$ indicating the distributions are identical and $B=0$ indicating that the PDFs are perfectly distinguishable for any value $t$.  Effectively, $B({\Pr}_1(t),{\Pr}_2(t))$ replaces the overlap integral convention typically used for quantum states. The utility of this adjustment stems from the fact that while overlap integrals are defined for square-normalized distributions, all PDFs are integral-normalized. Direct comparison of the sampled marginal with the state marginal provides a direct quantification of measurement quality without requiring full state tomography, and thus makes $B({\Pr}_1(t),{\Pr}_2(t))$ well suited for homodyne detection.

\subsection{Modal Mismatch}

The presence of measurement errors can directly impact the matching of temporal modes, which translates into a statistical mixture of the true quantum state with vacuum. In reconstructing the marginal, a measurement mode $\tilde{f}(t)$ must be selected. Ideally, $\tilde{f}(t)$ is equal to the principal temporal mode $f(t)$. However, due to chromatic dispersion, spectral filtering, or measurement error, $\tilde{f}(t)$ will be modified from the theoretical expectation, and lead to an imperfect selection of $\tilde{f}(t)$. The efficiency of the homodyne measurement can then be determined by the overlap
\begin{equation} \label{eq:overlap}
    \eta=\langle f\vert\tilde{f}\rangle
    =\int f^*(t)\tilde{f}(t) dt.
\end{equation}
The resulting measurement samples the state in $\tilde f(t)$ with likelihood $\eta$, and similarly samples the states in each orthogonal mode with probability defined by equation (\ref{eq:overlap}) for the entire basis. We consider a single-mode state, such that all orthogonal modes contain the vacuum and the measured marginal can be modeled as \begin{equation}
\hat q_{\theta}^{(\tilde f)}
= \eta\, \hat q_{\theta}^{(f)}
+ \sqrt{1-\eta^2}\, \hat q_{\theta}^{(\mathrm{vac})}.
\end{equation} 
Sampling from this marginal results in a statistical mixture of the quantum state marginal and vacuum marginal, with the proportion of each determined by the degree of modal mismatch.  
It is important to note that even complete temporal mode mismatch will not lead to $B=0$, but will instead compare the quantum state marginal to the vacuum state marginal. A low photon number coherent state $\ket{a}$, $\alpha\ll1$ will have a significant overlap with the vacuum state whereas a coherent state with $\alpha\gg1$ will have little overlap. There is therefore state-dependence to the interpretation of $B$, which is an important consideration when identifying the principal mode using experimental data. Our algorithm hence allows us to explore these errors, as shown in Figure~\ref{fig:multi_jitter}(a), across various states to inform the expected sensitivity to modal mismatch. 

\subsection{Timing Jitter}

Uncertainty in the arrival time of the quantum state leads to a random temporal displacement of the signal mode corresponding to each homodyne trace. For an ideal temporal mode $f(t)$, the timing jitter resulting in a displaced mode can be defined as
\begin{equation}
f(t) \rightarrow f(t-\tau).
\end{equation}
The timing offset $\tau$ is represented by a zero-mean Gaussian distribution with standard deviation $\sigma$ \cite{kumar2012versatile}.

\begin{figure}
    \centering
    \includegraphics[width=\linewidth]{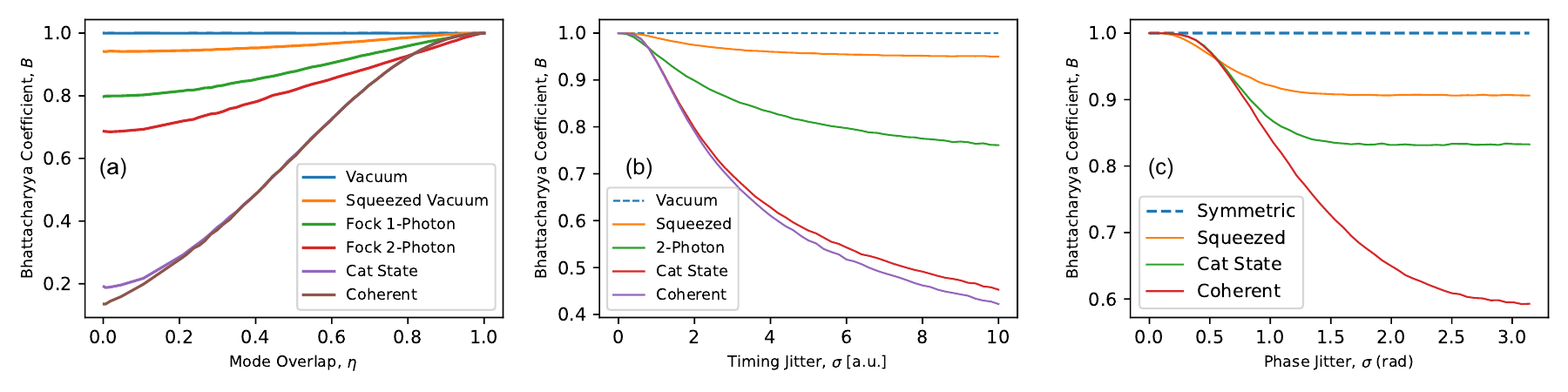}
    \caption{Similarity between theoretical marginal and reconstructed marginal subject to different measurement errors. (a) \textit{Mode overlap error}: A temporal mode $f(t)$ is reconstructed for various translations $f(t-t_0)$ resulting in different mode overlap. (b) \textit{Timing jitter}: A temporal mode $f(t)$ is subject to random temporal shifts $f(t-\tau)$ for $\tau$ drawn from a normal distribution with standard deviation $\sigma$, averaged over $10^5$ realizations. (c) \textit{Phase jitter}: The mode is always reconstructed in the correct temporal mode, but subjected to a random rotation in phase space. Note that symmetric states such as a Fock state are insensitive to these phase errors.}
    \label{fig:multi_jitter}
\end{figure}

To simulate this effect, we generate sets of photocurrent traces with timing offsets $\tau$ drawn from the Gaussian distribution. Quadrature marginals are reconstructed by integrating each trace against a fixed reference mode, and then they are compared to the ideal marginal using the Bhattacharyya coefficient. Figure~\ref{fig:multi_jitter}(b) shows the resulting $B$ coefficient as a function of the timing jitter width $\sigma$ for several representative quantum states. 

\begin{figure}
    \centering
    \includegraphics[width=\linewidth]{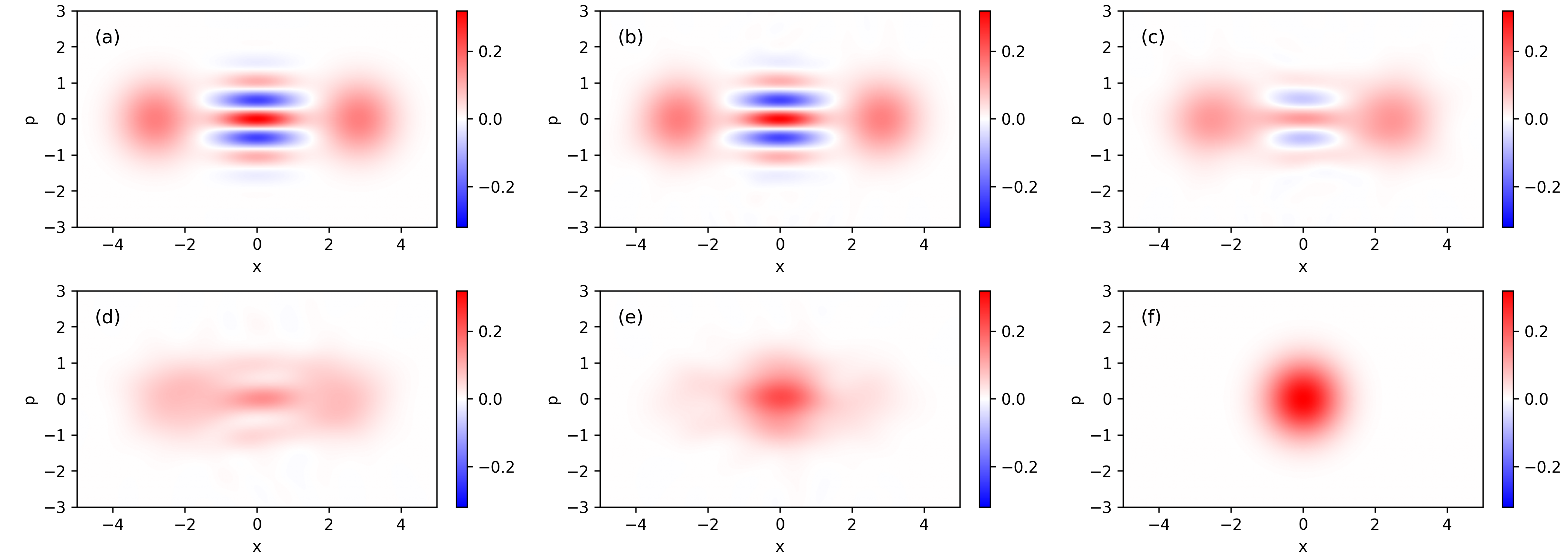}
    \caption{Wigner functions of maximum likelihood quantum state reconstruction subject to measurement error. The reconstructed Wigner function approaches the vacuum state as errors become large. (a) The original cat state in a Gaussian mode with standard deviation $\sigma$, to generate the photocurrent traces. (b), (c), (d), (e) The reconstructed Wigner function subject to a timing jitter standard deviation of $0$, $\sigma$, $2\sigma$ and $5\sigma$ respectively. (f) The theoretical vacuum state Wigner function.}
    \label{fig:reconstructed_W_functions}
\end{figure}

Furthermore, to demonstrate the effect of timing jitter, we reconstruct a cat state subject to a range of timing jitter standard deviations. There are a number of approaches to Wigner function reconstruction, like the inverse Radon transformation \cite{RadonJohann1986} or Maximum Likelihood (ML) methods \cite{lvovsky2004MaxLik}. We chose to use the ML method due to its insensitivity to numerical artifacts, and its widespread experimental adoption. Figure \ref{fig:reconstructed_W_functions} shows how the reconstructed Wigner function tends towards the vacuum state as errors become large.

\subsection{Phase Jitter}
Phase jitter corresponds to phase variation in the quadrature angle
\begin{equation}
\theta \rightarrow \theta + \delta\theta.
\end{equation}

This effect can arise from path length shifts of the LO, such as mirror vibration or thermal fluctuations in air. Due to the stochastic nature of these variations, we model the reconstructed marginal as an average over realizations of phase angles. Specifically we model phase fluctuations by a Gaussian probability density $p(\delta\theta)$. By the law of the total probability, the reconstructed marginal can be expressed as,
\begin{equation}
\Pr_{\mathrm{rec}}(q)
= \int d(\delta\theta)\,
p(\delta\theta)\,
\Pr_\rho(q_{\theta+\delta\theta}).
\label{eq:phase_avg}
\end{equation}

This expression corresponds to an angular averaging of Wigner function projections, rather than a convolution in quadrature space. Unlike timing jitter, phase jitter does not necessarily induce temporal mode mismatch and can therefore act independently of mode overlap. For phase invariant states such as Fock states,
\begin{equation}
\Pr(q_{\theta+\delta\theta}) = \Pr(q_\theta)
\quad \forall\, \delta\theta,
\end{equation}
and thus phase jitter has no effect. For asymmetric states, however, phase jitter can dominate the error observed in experiment. Figure \ref{fig:multi_jitter}(c) shows the measurement as a function of similarity for several quantum states.


\section{Conclusions}
\label{sec:conc}

We have presented a conceptual and algorithmic framework for the simulation of continuous homodyne detection of time dependent quantum states of light. In this picture, the homodyne photocurrent is naturally expressed in the detector's ``time-bin basis'' whereas the quantum state is most conveniently expressed in its principal basis, defined by the process from which it is generated. We can consider homodyne detection as a transformation from the one basis to another, and thus the recorded photocurrent is a projection of the states natural basis to the detector's basis at each instance of time.

Using this picture, we presented an algorithm for efficiently simulating photocurrent which results from balanced homodyne detection. Specifically, starting with a quantum state in a given temporal mode, we considered a measurement of the quadrature of this state using an $N-$sample trace from a balanced homodyne photodetector. We used efficient numerical factorization to represent this state as part of a $N-$dimensional vector space, with all modes but the principal quantum mode taken as samples of the vacuum state. This represents a realistic photocurrent trace when measured continuously.

We then demonstrated the utility of our algorithm by using it to simulate several common experimental errors in quantum state tomography and quadrature measurement. Specifically, we analyzed the effect of timing jitter, phase noise and modal mismatch due to group velocity dispersion. Although we analyzed a single mode quantum state in a vacuum background, our method can directly generalize to multimode states in an arbitrary background, such as a thermal state. Open-source code used to perform our analysis is available in a GitHub repository \cite{github2026}. We hope that the work presented herein will be useful to practitioners seeking high fidelity reconstruction of the quantum state and would like to determine error bounds.

\vspace{-20pt}
\acknowledgments{We acknowledge support of the Natural Sciences and Engineering Research Council of Canada (NSERC) (grant RGPIN-2021-03289).}




\bibliographystyle{unsrt}
\bibliography{simhomorefs}

\end{document}